\documentclass[aps,prl,preprint,superscriptaddress,amsmath,amssymb]{revtex4}
\usepackage{graphicx}
\begin{document}
\title{Ionization of Xe in intense, two-cycle laser fields: Dependence on carrier-envelope phase}

\author{Parinda Vasa}
\affiliation{Department of Physics, Indian Institute of Technology Bombay, Powai, Mumbai 400 076, India}
\author{Aditya K. Dharmadhikari}
\affiliation{Tata Institute of Fundamental Research, 1 Homi Bhabha Road, Mumbai 400 005, India}
\author{Jayashree A. Dharmadhikari}
\affiliation{Department of Atomic and Molecular Physics, Manipal University, Manipal 576 104, India}
\author{Deepak Mathur}
\email{atmol1@tifr.res.in}
\affiliation{Tata Institute of Fundamental Research, 1 Homi Bhabha Road, Mumbai 400 005, India}
\affiliation{Department of Atomic and Molecular Physics, Manipal University, Manipal 576 104, India}
\date{\today}

\begin{abstract}
We report on tunnel ionization of Xe by 2-cycle, intense, infrared laser pulses and its dependence on carrier-envelope-phase (CEP). At low values of optical field ($E$), the ionization yield is maximum for cos-like pulses with the dependence becoming stronger for higher charge states. At higher $E$-values, the CEP dependence either washes out or flips. A simple phenomenological model is developed that predicts and confirms the observed results. CEP effects are seen to persist for 8-cycle pulses. Unexpectedly, electron rescattering plays an unimportant role in the observed CEP dependence. Our results provide fresh perspectives in ultrafast, strong-field ionization dynamics of multi-electron systems that lie at the core of attosecond science.
\end{abstract}

\maketitle
Probing atomic and molecular dynamics in the strong field regime by means of few-cycle laser pulses with stabilized carrier-envelope-phase (CEP) has begun to provide a fillip to attosecond science \cite{Chang}. Within a single laser pulse, CEP is a measure of the temporal offset between the maximum of the optical cycle and that of the pulse envelope. CEP-stabilized laser pulses present opportunities to control, within a fixed intensity pulse, the instantaneous magnitude of the time-dependent optical field. This, in turn, permits control to be exercised not only on the moment of ``birth" of an electron's wave packet as it tunnels through the field-distorted atomic Coulomb potential but also on its subsequent motion in the oscillating optical field. Theoretical treatments of the intensity-dependent CEP response of an atom in an ultrashort optical field remain a challenge for it necessitates knowledge of the population probabilities and phases of different Floquet states. To this end, it becomes mandatory to numerically solve the time-dependent Schr\"odinger equation in which the field-free Hamiltonian is supplemented by a term that accounts for the interaction with the optical field \cite{Brabec,Shvetsov}. It is, therefore, understandable that theoretical endeavors in this area of strong field science continue to face very formidable hurdles. We present here experimental results on how field-induced single and multiple ionization of Xe atoms by 5 fs pulses of 800 nm light depends on CEP. The peak intensity of our 2-cycle pulses (2-5 PW cm$^{-2}$) ensures that the ionization dynamics occur in the tunneling regime. It is known that the CEP of the driving laser pulse significantly influences high harmonic generation (HHG) \cite{Ishii}, the generation of isolated attosecond pulses \cite{Heyl}, multiphoton ionization \cite{Nakajima}, and above threshold ionization (ATI) \cite{Schumacher}. Indeed, it is these processes that have attracted overwhelming contemporary attention vis-\`a-vis their dependence on CEP \cite{Niikura}. These phenomena are complex as they involve multiple steps. There remains an acute paucity of information on the CEP dependence of the quantum mechanical aspects of the {\em primary} strong-field process - the field-induced distortion of the atomic potential. It is this facet that we explore in the present study as it provides distinct advantages over exploring CEP dependence of complex processes like HHG.

HHG is a direct consequence of the electron rescattering process \cite{kuchiev}. Within a 2-cycle pulse, there are multiple instances when the probability of ionization followed by rescattering of the ionized electron by the ionic core is maximum. The HHG spectrum is a superposition of the photon emission at all such instances \cite{Ishii}, making it difficult to correlate photon emission with electron dynamics. As we show in the following, the maximum probability for photon emission does not coincide with the peak of the optical field. Thus, the manifestation of CEP dependence is expected to be complicated in the HHG process \cite{Ishii}. In contrast, the quantum mechanical process of tunnel ionization is a direct manifestation of the field-induced distortion of the atomic potential; its probability is strongly field dependent (it maximizes at the peak of the optical field), making this channel extremely CEP-sensitive. Here, we focus attention on tunnel ionization of a multi-electron atom, Xe which, through generation of high harmonics, has found widespread utility in attosecond science. We have selected intensity values that ensure tunnel ionization of Xe: the optical field is kept small enough to ensure that the barrier-suppression regime is not accessed. This has a bearing on the modeling that we develop to rationalize our observations. Our results show stronger CEP dependence at lower values of optical field ($E$) and for higher charge states. At higher intensities ($\sim$7 PW cm$^{-2}$) sequential ionization can occur: lower charge states formed at the rising edge of the pulse are further ionized as $E$ approaches its peak value. In such cases, the CEP-dependence of the lower charge state is either washed out or it flips. Our phenomenological model rationalizes these results and also predicts CEP-dependent behavior when longer pulses (22 fs) are used. Our experiments confirm these predictions. Conventional wisdom, for a long time, was that CEP effects only manifest themselves when the pulse duration is nearly as short as the optical period. However, theoretical as well as experimental evidence has begun to offer hints that CEP affects the strong field dynamics of atoms even for longer duration pulses \cite{long1,long2}, making CEP effects of more widespread significance than originally thought. We have also probed the role of electron rescattering \cite{kuchiev} in few-cycle ionization dynamics and establish that electron kinetic energy values at times when the rescattering probability is maximum are only marginally affected by CEP. Unexpectedly, rescattering seems to be relatively unimportant in determining CEP-dependence because the rescattering probability is not the highest when there is maximum field distortion.

We used a Ti:sapphire oscillator (75 MHz repetition rate) whose pulses were amplified by a 4-pass amplifier, then stretched to $\sim200$ ps, before being directed through an acousto-optic dispersive filter that controlled pulse shape and duration. The resulting output passed, via an electro-optical modulator (which down-converted to 1 kHz repetition rate), to a 5-pass amplifier and compressor to yield a 22 fs pulse which was further compressed to 5 fs by a 1-m-long Ne-filled hollow fiber and a set of chirped dielectric mirrors. For CEP stabilization we used a fast-loop in the oscillator and a slow-loop in the amplifier \cite{11}. Interferometric correlation measurements showed that, typically, the phase jitter obtained in the course of our experiments was $<$60 mrad for 5 fs pulses. Our $f-2f$ interferometer operated at 1 kHz spectrometer acquisition rate, with 920 $\mu$s integration time and 84 ms loop cycle. The stability of laser energy with and without CEP stabilization was 0.4\% and 1.7\% rms, respectively. Linearly polarized pulses were transmitted through a thin (300 $\mu$m) fused-silica window \cite{12} to an ultrahigh vacuum chamber (10$^{-11}$ Torr) in which a Xe atomic beam intersected our laser beam focused with a 5 cm curved mirror to 7 $\mu$m ($1/e^{2}$ width) \cite{13}. Xe ionization was monitored using a linear (20 cm) time-of-flight spectrometer interfaced to a data acquisition system (1 kHz) on a segmented-mode 2.5 GHz oscilloscope. Peak laser intensity at the focal spot was determined with reference to the appearance intensities of Xe$^{2+}$ ($\sim$2 PW cm$^{-2}$) and Xe$^{3+}$ ($\sim$2.5 PW cm$^{-2}$) \cite{SM}. These intensity values correspond to Keldysh parameters 0.1-0.5 that ensure the tunnelling ionization regime.

\begin{figure}
\includegraphics[width=8.0cm,clip]{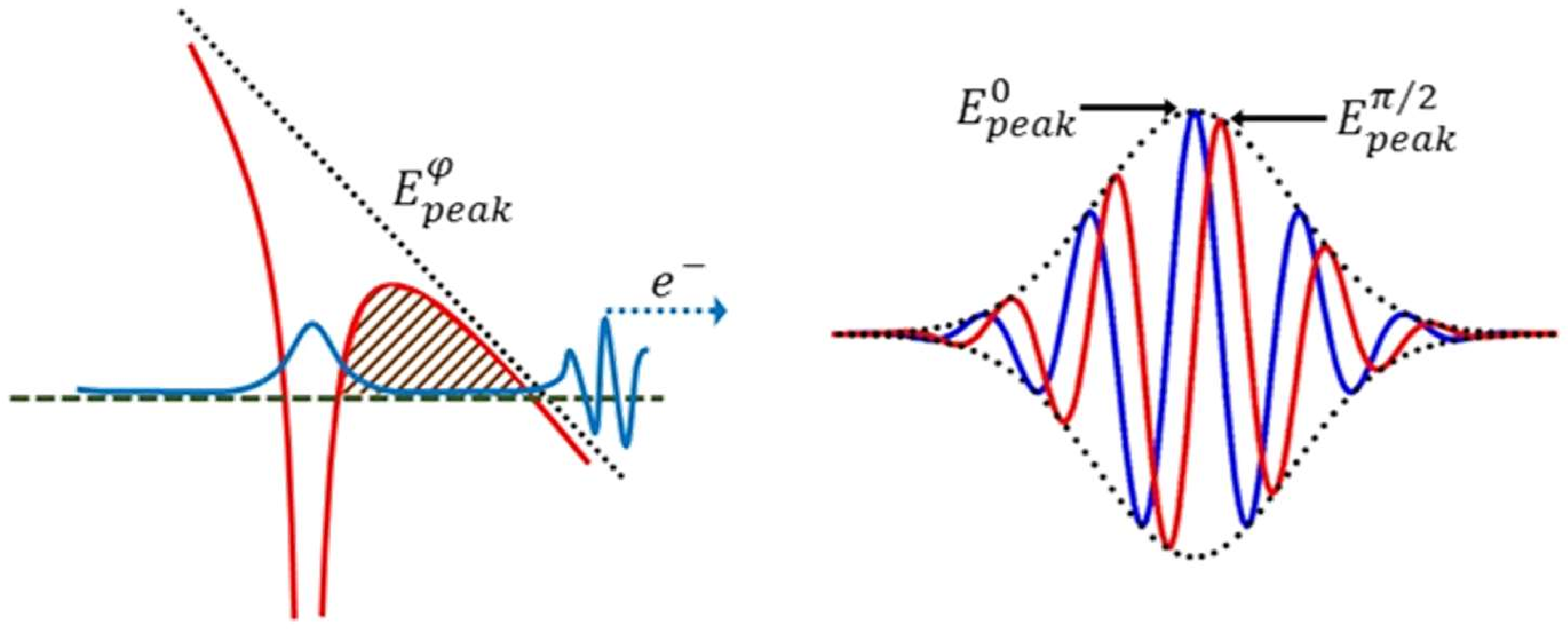}%
\caption{\label{fig1} {\footnotesize (color online).
Left panel: Schematic representation of our experimental situation. Tunnel ionization of Xe occurs through the shaded area. Right panel: Temporal profiles of the optical field in CEP = 0 (cos-like) and CEP = $\pi$/2 (sine-like) pulses.
}}
\end{figure}

Figure 1 characterizes our experimental situation wherein tunneling is through the shaded area with a probability that varies exponentially with the barrier area. Hence, the regime accessed in our experiments is characterized by an exponential dependence of the ionization rate on the transient distortion of the atomic potential which, in turn, is governed by the instantaneous value of $E$. The oft-used ADK (Ammosov-Delone-Keldysh) formulation \cite{ADK} yields the tunneling probability ($W$):

\begin{equation}
{W=\frac{Q^{2}}{2n^{*2}}\sqrt{\frac{3En^{*3}}{\pi Q^{3}}}~~{\rm exp}\left({-\frac{2Q^{3}}{3n^{*3}E}}\right)},
\end{equation}

\noindent where $Q$ is the ion charge and $n^{*}$ is the effective principal quantum number. As eqn. (1) yields the probability of creating different charge states from neutral Xe, it predicts extremely small probability of creating higher charge states and is of limited utility in the presence of sequential ionization. As already noted, the ionization probability scales exponentially with $E$ and, consequently, it will change considerably with even a small variation in $E$. In the case of our two-cycle pulse, even marginal alteration of the CEP value has a discernible effect on the amplitude of $E$. Thus, scanning the CEP is analogous to changing the area of the potential barrier. Figure 1 shows that a cos-like pulse (CEP=$\phi=0$) has ~5\% higher peak amplitude compared to a sine-like pulse ($\phi=\pi$/2) of the same peak intensity. Although $E$ changes only marginally in these two pulses, there is substantial change in the tunneling rate: a $\phi=0$ pulse will induce as much as $\sim$20\% more ionization than a $\phi=\pi/2$ pulse of the same peak intensity. Since all other experimental conditions are maintained identical during the course of measurements, we take the measured ion-yield to be a direct manifestation of ionization rate [eqn. (1)]. The ADK formulation predicts stronger CEP-dependence at lower $E$-values and higher charge states (Xe$^{2+,3+}$), in consonance with our measured data (Fig. 2). We obtain ionization yields for different charges, at a given CEP value, from the same time-of-flight spectrum \cite{SM}. Our measurements reveal a marked CEP-dependence for charge states greater than $1+$, with maximum ion-yield obtained at $\phi=0$. At higher $E$-values, the distortion in the potential becomes large enough to approach the barrier suppression regime, at which the ionization rate scales linearly with the field amplitude (Xe$^+$ data, Fig. 2). Here, the variation in ion-yield over a half-cycle is only $1\%$, which is well within the power and CEP stability of our laser system. For higher charge states, a systematic increase in variation is observed [as high as $22\%$ variation, Fig. 2(b)-2(d)]. It is important to note that the observed CEP-dependence gets ``washed out" when jitter in CEP stabilization exceeds $\sim$150 mrad (see Fig. S2 in Supplemental Material and \cite{Lezius}).

\begin{figure}
\includegraphics[width=8.0cm,clip]{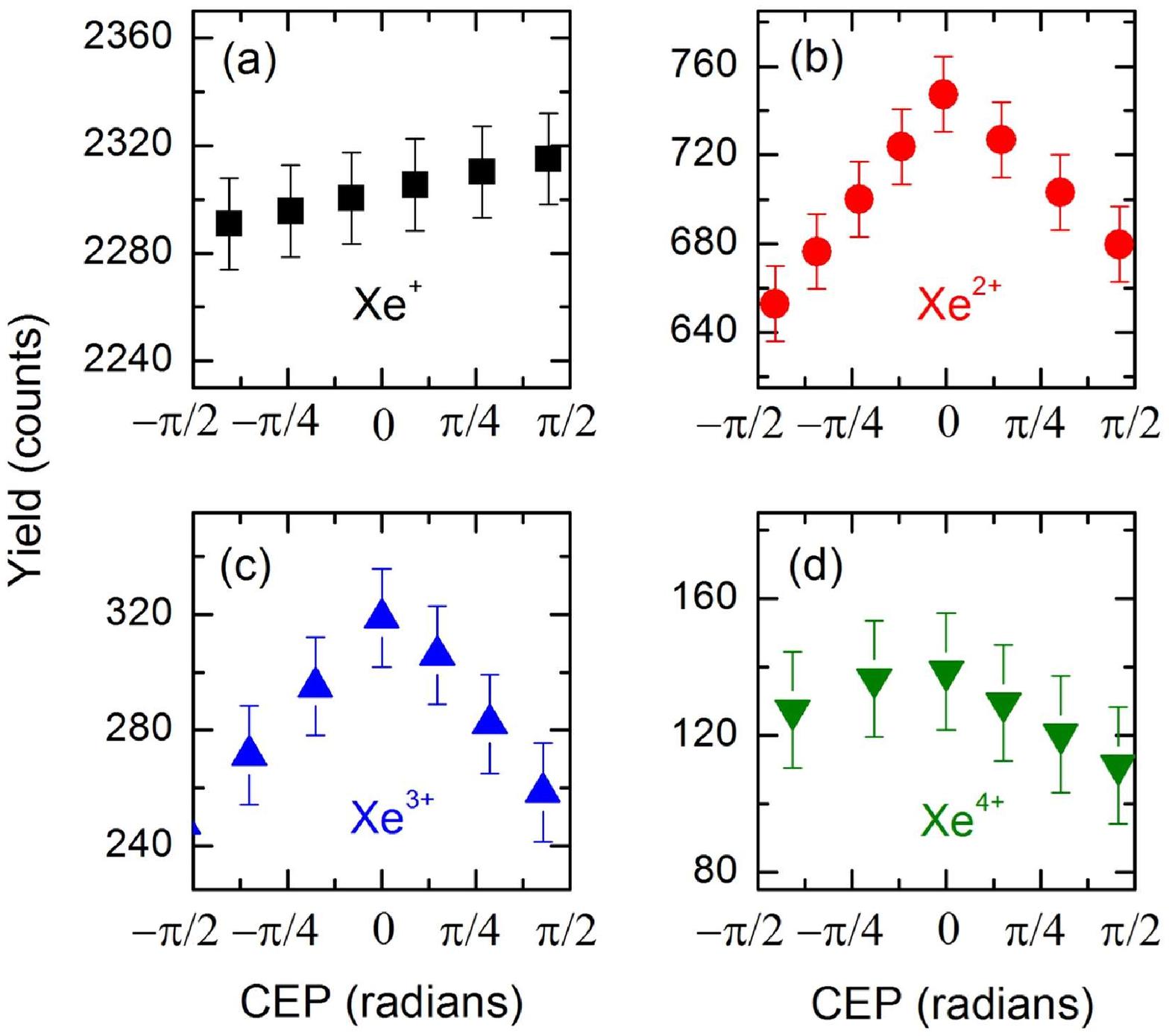}%
\caption{\label{fig2} {\footnotesize (color online).
CEP-dependent ionization yields of (a) Xe$^{+}$, (b) Xe$^{2+}$, (c) Xe$^{3+}$, and (d) Xe$^{4+}$  obtained using 5 fs pulses with peak intensity of  $\sim$2.5 PW cm$^{-2}$. The ratio of ion-yield at $\phi=0$ and $\phi=\pi/2$ is 0.99 for Xe$^{+}$, 0.9 for Xe$^{2+}$, 0.87 for Xe$^{3+}$, and 0.78 for Xe$^{4+}$. Note the near independence on CEP of ion-yields for low charge states, and increasing CEP-dependence for higher charge states.
}}
\end{figure}

At sufficiently high intensity, there is the possibility of significant sequential ionization (SI) setting in. The wings of the two-cycle pulse may then be intense enough to create a lower Xe charge state which is then further ionized as $E$ increases. In such cases, the CEP-dependence of the lower charge states may be totally washed out or may even flip, as shown in Fig. S3 in Supplemental Material and in an earlier calculation \cite{Shvetsov}. Since the number of higher charge states detected depends on the number of lower charge states created within the same pulse, the probability of generating the former is maximum for a $\phi=0$ pulse. However, this depletes the number of low charge ions that are detected, thus flipping the CEP-dependence (maximum at $\phi=\pi$/2) of the lower charge state (Fig. 3).

\begin{figure}
\includegraphics[width=8.0cm,clip]{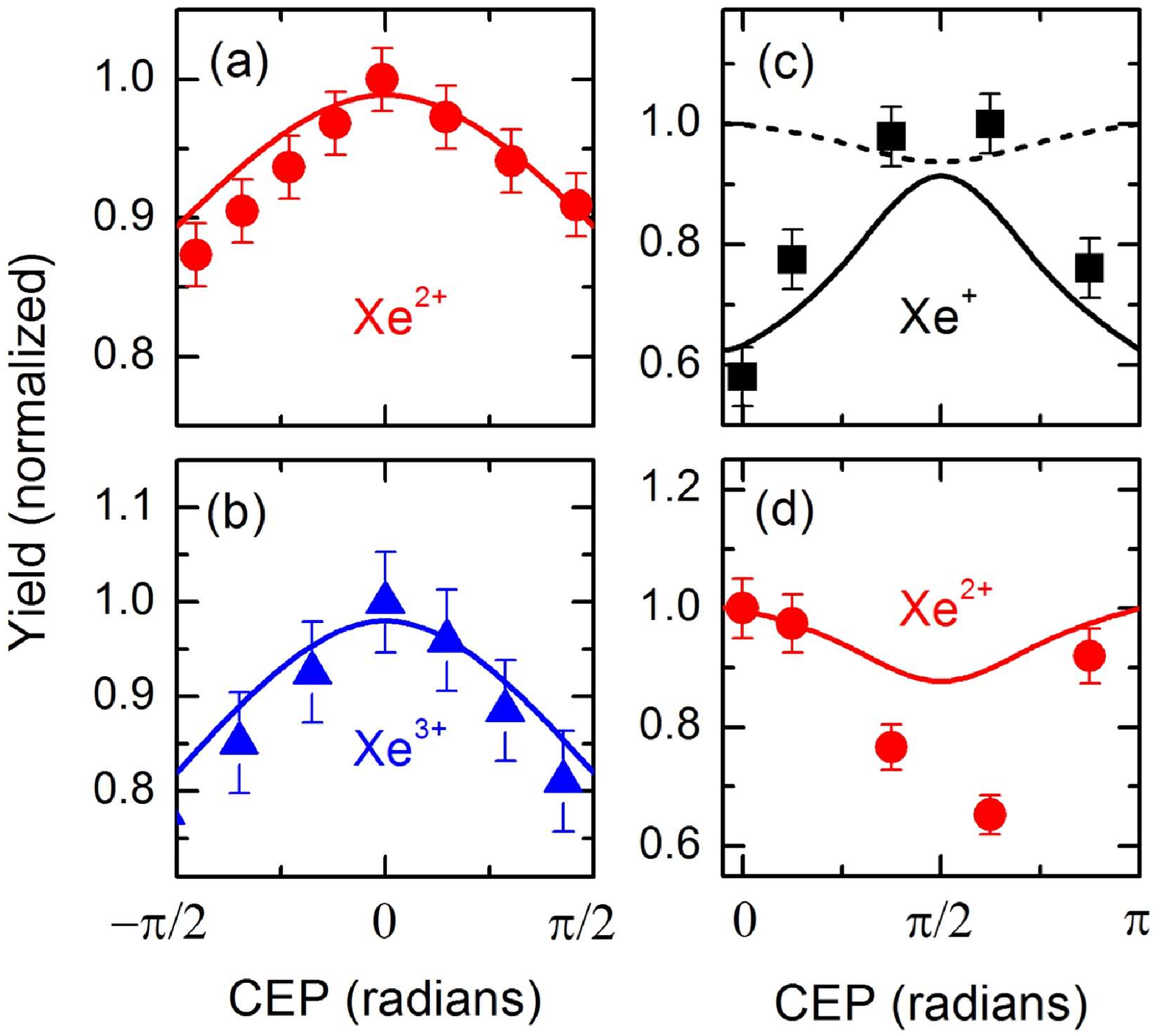}%
\caption{\label{fig3} {\footnotesize (color online).
Comparison of measured and calculated ionization yields over half-CEP period of (a) Xe$^{2+}$ and (b) Xe$^{3+}$ without substantial sequential ionization (SI) as in Fig. (2). (c,d) show the corresponding comparison of yields of Xe$^{+}$ ($W_{\phi}^{+}$, solid line) and Xe$^{2+}$ ($Z_{\phi}$, solid line) in the presence significant of SI (higher peak intensity $\sim5$ PW cm-2). (c) also shows the calculated yield of Xe$^{+}$ in the absence of SI ($W_{\phi}^{-}$, dashed line). Note how incorporation of SI flips the CEP dependence in (c).
}}
\end{figure}

We have developed a model to account for these observations. Taking $N_1$ to be the number of Xe$^+$ ions formed in the absence of SI, $N_1=\alpha_1N$, where $N$ is the total number of Xe-atoms, and $\alpha_1$ is the probability of ionizing Xe$^+$. Similarly, if $N_2$ is the number of Xe$^{2+}$ ions formed in the presence of SI, $N_2=\alpha_2N_1$. The CEP dependence of Xe$^{+}$ (with respect to $\phi$=0) is then

\begin{equation}{
W_{\phi}^{-}=\frac{\alpha^{\phi}_{1}}{\alpha^{0}_{1}}
}\end{equation}
when SI is negligible and as
\begin{equation}{
W_{\phi}^{+}=\frac{\alpha^{\phi}_{1}N-\alpha^{\phi}_{2}\alpha^{\phi}_{1}N}{\alpha^{0}_{1}N-\alpha^{0}_{2}\alpha^{0}_{1}N}=W_{\phi}^{-}\frac{1-\alpha_2^\phi}{1-\alpha_2^0}
}\end{equation}

when SI is significant. It is the factor $(1-\alpha_2^\phi)/(1-\alpha_2^0 )$ (maximum at $\phi=0$), that flips the CEP dependence of the lower charge state (maximum at $\phi=\pi$/2). For the higher charge state,

\begin{equation}
Z_{\phi} = {\alpha_2^\phi\over{\alpha_2^0}}.
\end{equation}

Accordingly, Fig. 3 depicts the CEP-dependence of Xe$^+$ and Xe$^{2+}$ being out-of-phase with each other, in consonance with our observations.

In order to explore whether CEP-dependent ionization is only of concern with few ($\sim$2)-cycle pulses we also made measurements using $\sim$8-cycle ($22$ fs) pulses of $\sim$7 PW cm$^{-2}$ intensity (Fig. 4). The yields of charge states $1+$, $2+$, and $3+$ appear to to be CEP-independent. CEP-dependence manifests itself for charge states $>3+$ and is strongest for Xe$^{6+}$. The rising edge of the longer pulse provides sufficient field to induce ionization to charge states $1+$, $2+$, and $3+$; it is only in the vicinity of the peak of the pulse that the field is strong enough for manifestation of CEP dependence for charge states $>3+$, as expected from our model.

\begin{figure}
\includegraphics[width=8.0cm,clip]{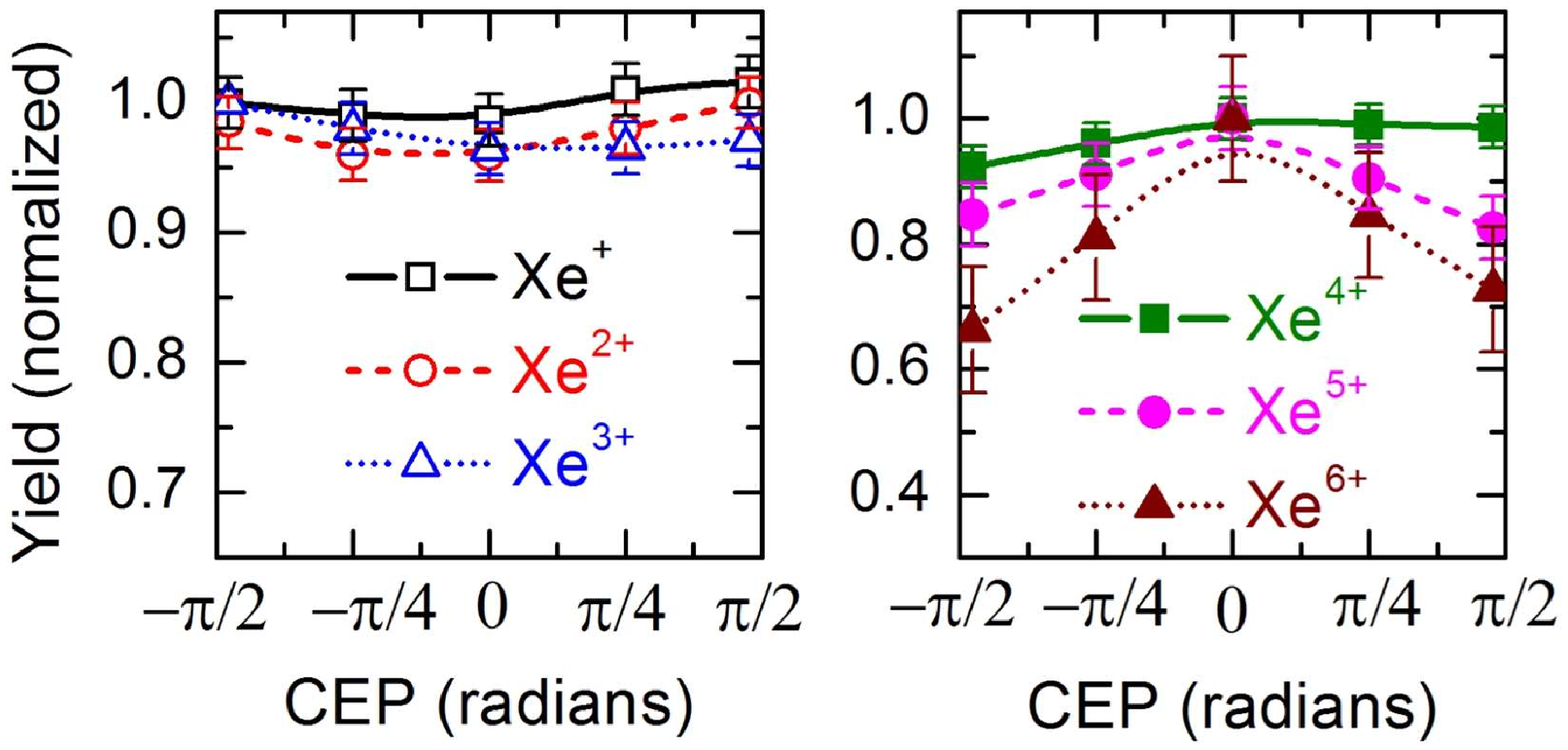}%
\caption{\label{fig4} {\footnotesize (color online).
Ionization yields of Xe$^{+}$, Xe$^{2+}$, and Xe$^{3+}$ (left panel), and Xe$^{4+}$, Xe$^{5+}$, and Xe$^{6+}$ (right panel) measured using 22 fs ($\sim$8-cycles) pulses. Note the CEP independence of the three lowest charge states (see text). Lines correspond to a spline fit to the data.}}
\end{figure}

We also investigated the role of electron rescattering \cite{kuchiev} wherein the ionized electron is accelerated by the optical field and upon sign reversal of $E$, it accelerates back towards the residual ion. Using classical equations of motion, we map electron trajectories (distance from the ionic-core) for $\phi=0$,$\pi$/2 [Fig. 5 (a,b)]. Typical electron trajectories corresponding to three ionization times [Fig. 5(c)] show that rescattering occurs only when the electron is generated close to ionization times denoted by black dashed lines [Fig. 5 (a,b)]. In fact, the rescattered electron possessing the highest energy is generated 0.3 radian after the peak of $E$ [circles in Fig. 5(d)]. These electrons re-encounter the ionic core at a later time, as indicated by arrows in Fig 5(d). Such instances do not correspond to the peak of $E$ (where the probability of tunnel ionization maximizes). This temporal mismatch suggests that re-scattering is not the prime driver of sequential ionization in the tunnelling regime.

Furthermore, since the ponderomotive energy varies as $E^2$ and there is only a $5\%$ difference in the peak value between $\phi=0$ and $\pi$/2 pulses, the maximum kinetic energies at the time of rescattering for different CEP values are comparable for 5 fs pulses. In contrast, the exponential dependence of tunnelling probability on $E$ is expected to result in much stronger CEP dependence even for a mere $5\%$ difference in the peak-field amplitude. Hence, we are led to an unexpected conclusion: although rescattering is known to dominate HHG \cite{Ishii} and multiphoton double ionization of Ar \cite{Johnson}, it plays a minor role in determining the CEP-dependence, particularly for higher charges. This is because the rescattering probability does not maximize at the peak of the field amplitude. At sufficiently high intensity, in the barrier suppression regime, tunnel ionization is no longer an effective and rescattering could become important.
	
Our results with 22 fs pulses also support this conjecture. For these pulses, there will be many more instances when rescattering could occur, completely washing out the CEP dependence. However, strong CEP dependence does persist for higher charges (Fig. 4) which, due to the non-adiabatic nature of sequential ionization, are generated only near the peak of the pulse. Therefore, for charge states $>3+$, our eight-cycle pulse is essentially like a two-cycle pulse vis-\`a-vis tunnel ionization, preserving the CEP dependence.

\begin{figure}
\includegraphics[width=8.0cm,clip]{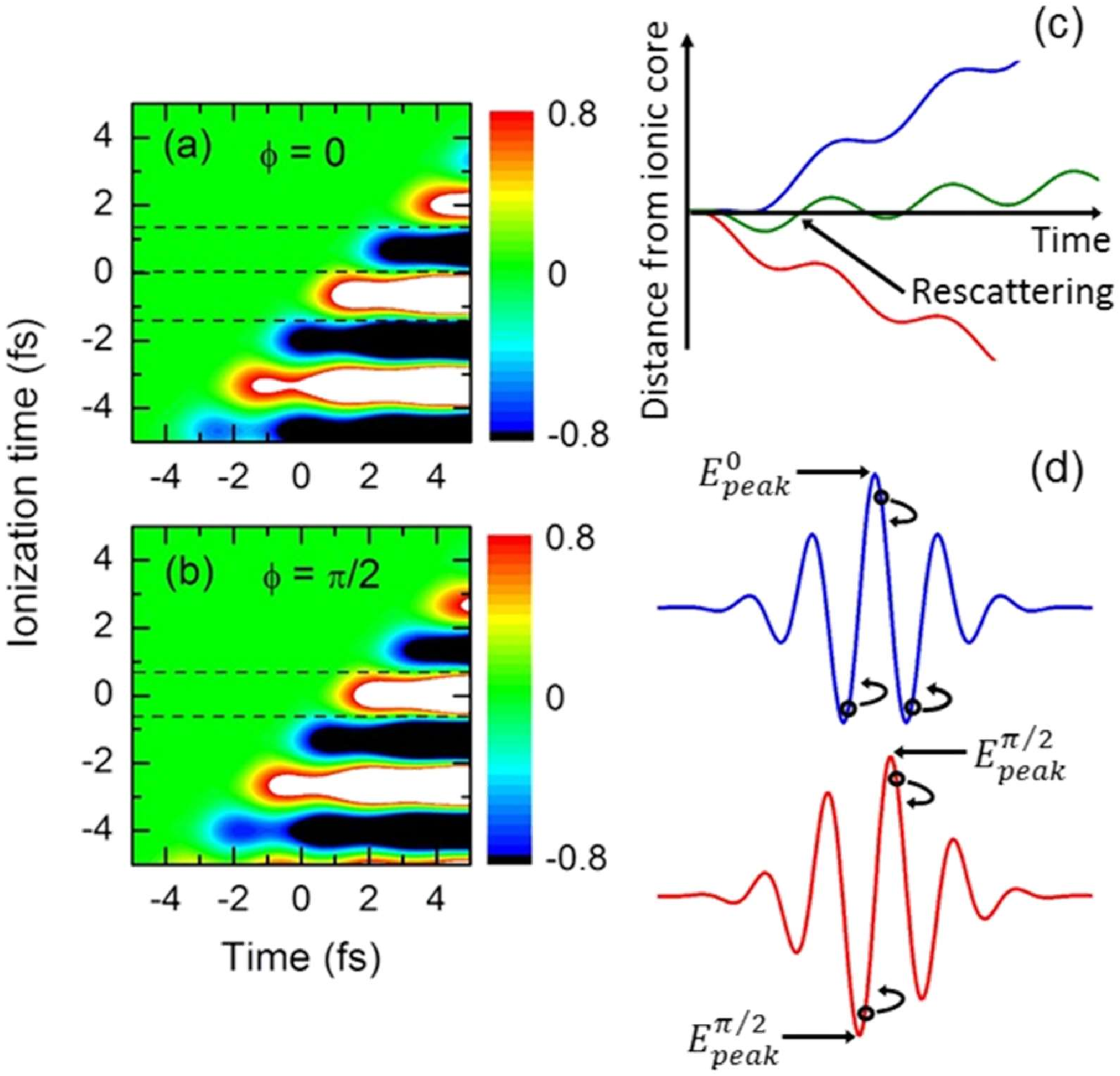}%
\caption{\label{fig5} {\footnotesize (color online).
Temporal evolution of (a,b) electron trajectories at CEP = 0,$\pi$/2. The black dashed lines mark the ionization times at which the probability of re-encountering the ionic-core (rescattering) is maximum; (c) typical electron trajectories for different ionization times. Only some of the trajectories mapped in (a,b) result in rescattering; (d) Schematic depicting instances near the peak of the pulse ($\phi$ = 0,$\pi$/2) corresponding to ionization times with highest rescattering probability (circles). These generated electrons re-encounter the ionic core at a later time, denoted by arrows. Note the temporal mismatch between the peak of the $E$-field (with highest tunnelling probability) and rescattering (with maximum kinetic energy).
}}
\end{figure}

Our results on CEP-dependent ionization of Xe open new vistas in attosecond science in the truly quantum mechanical regime. For instance, higher intensity CEP-stabilized pulses might enable control to be exercised on electronic motion in the inner orbitals, enabling new classes of experiments on heavy atoms in which such electrons are relativistic. Also, there is little insight about the interplay of inner- and outer-shell electrons in multielectron systems. There has been inconclusive debate on how (and if) external fields may be shielded from electrons in inner orbitals \cite{24}; proper descriptions of the dynamics in multielectron systems remain intractable. Our experiments should aid in testing the efficacies of future theoretical developments in this direction. In the case of molecules made up of heavy atoms, dipole and polarizability corrections need to be incorporated into existing strong-field theories; such corrections are, of course, CEP-dependent and we anticipate that our results will stimulate further work.

Financial support from the Department of Science and Technology is acknowledged by J. A. D. (Women Scientists Scheme) and by D. M. (J. C. Bose National Fellowship). We are grateful to Krithika Dota for assistance in the early stages of this work.

\end{document}